\documentclass[11pt,a4paper]{article}
\usepackage{fullpage}
\usepackage{amsmath, amsthm, amsfonts,  amssymb, latexsym}
\usepackage[colorlinks, citecolor=Red, linkcolor=Blue, urlcolor=blue]{hyperref}
\usepackage{caption}
\usepackage[dvipsnames]{xcolor}
\usepackage{tikz}
\usepackage{pgfplots}
\usetikzlibrary{arrows,arrows.meta, decorations.pathmorphing, decorations.markings, decorations.pathreplacing,intersections, pgfplots.fillbetween,patterns,shadings}
\usepackage{enumerate}

\newcommand{\rhs}{r.h.s.\ }
\newcommand{\lhs}{l.h.s.\ }
\newcommand{\wrt}{w.r.t.\ }
\newcommand{\cf}{cf.\ }

\makeatletter
\DeclareRobustCommand{\ibar}{\mathord{%
  \text{$\m@th\mkern-2mu\raisebox{-0.7ex}[0pt][0pt]{$\mathchar'26$}\mkern-7mu i$}%
}}
\makeatother

\newcommand{\bra}[1]{\langle #1 \rvert}
\newcommand{\ket}[1]{\lvert #1 \rangle}

\newcommand{\ud}{\mathrm{d}}
\newcommand{\del}{\partial}

\newcommand{\R}{\mathbb{R}}

\newcommand{\skal}[2]{\langle #1 , #2 \rangle}

\newcommand{\cL}{\mathcal{L}}

\newcommand{\eps}{\varepsilon}

\newcommand{\ret}{{\mathrm{ret}}}
\newcommand{\adv}{{\mathrm{adv}}}

\newcommand{\vp}{{\varphi}}

\newcommand{\nn}{\nonumber}
\newcommand{\beq}{\begin{equation}}
\newcommand{\eeq}{\end{equation}}

\newcommand{\defeq}{\mathrel{:=}}

\DeclareMathOperator{\sech}{sech}

\DeclareMathOperator{\Imag}{Im}

\newcommand{\cO}{{\mathcal{O}}}

\newcommand{\kink}{{\mathrm{kink}}}
\newcommand{\soliton}{{\mathrm{soliton}}}
\newcommand{\class}{{\mathrm{class}}}
\newcommand{\semiclass}{{\mathrm{semiclass}}}
\newcommand{\ps}{{\mathrm{ps}}}
\newcommand{\vac}{{\mathrm{vac}}}
\newcommand{\sG}{{\mathrm{sG}}}

\begin{document}

\title{The semiclassical energy density of kinks and solitons}
\author{Maria Angeles Alberti Martin\thanks{ma51gime@studserv.uni-leipzig.de}, Robert Schlesier\thanks{rs24fago@studserv.uni-leipzig.de} \ and Jochen Zahn\thanks{jochen.zahn@itp.uni-leipzig.de} \\ Institut f\"ur Theoretische Physik, Universit\"at Leipzig\\ Br\"uderstr.\ 16, 04103 Leipzig, Germany}

\date{\today}

\maketitle

\begin{abstract}
We compute semiclassical corrections to the energy density of kinks in $\phi^4$ theory and of solitons in the sine-Gordon model in $1+1$ dimensions, using local and covariant renormalization techniques from quantum field theory in curved spacetimes. For the semiclassical correction to the energy, we recover the known results. Our analysis highlights a subtlety in the definition of a conserved stress tensor for scalar field theories in $1+1$ dimension.
\end{abstract}

\section{Introduction}

Kink solutions of the $\phi^4$ model in $1+1$ dimensions are not only relevant for understanding properties of long-chain polymers, such as polyacetylene \cite{SuSchriefferHeeger1979}, but also have a long history as toy models for solitonic solutions in higher dimensions \cite{DashenEtAl}. In recent years, they have been reconsidered for example in the context of the Hamiltonian truncation method \cite{Coser:2014lla, RychkovVitale, Bajnok:2015bgw, Elias-Miro:2017tup} or Borel resummations \cite{Serone:2019szm}, in particular in relation to Chang duality \cite{Chang:1976ek}. 

Soliton solutions of the sine-Gordon model are intimately linked to the integrable structure of the theory, and there are strong indications that the WKB approximation for quantum corrections to their mass is exact \cite{Dashen:1975hd}. These results pass a number of non-trivial consistency checks \cite{Dashen:1975hd}, also in connection with the Coleman duality to the massive Thirring model \cite{Coleman:1974bu}. 

Here, we will reinvestigate the ``classical'' results of \cite{DashenEtAl, Dashen:1975hd} for semiclassical corrections to the energy of kinks and solitons from the point of view of modern local and covariant renormalization techniques. Specifically, we consider the $\phi^4$ and the sine-Gordon model in $1+1$ dimensions, specified by the potentials
\begin{align}
\label{eq:L}
 V_{\phi^4} & = \frac{\lambda}{4} \left( \phi^2 - \frac{m^2}{2 \lambda} \right)^2, &
 V_\sG & = \frac{m^4}{\lambda} \left[ 1- \cos \left( \frac{\sqrt{\lambda}}{m} \phi \right) \right].
\end{align}
These have vacuum solutions $\phi_\vac = \pm \frac{m}{\sqrt{2 \lambda}}$ and $\phi_\vac = 2 \pi n \frac{m}{\sqrt{\lambda}}$, respectively, and the parameters were chosen such that the fluctuations around these vacua have mass $m$ in both cases.\footnote{One recovers the conventions of \cite{DashenEtAl} by the replacement $m^2 \to 2 m^2$ in $V_{\phi^4}$.} A kink solution is a minimal energy stationary solution for the $\phi^4$ model that connects these two vacua, i.e., $\lim_{z \to \pm \infty} \phi_\kink(t, z) = \pm \frac{m}{\sqrt{2 \lambda}}$, whereas for the sine-Gordon model we consider solitons, i.e., stationary solutions connecting neighboring vacua. The semiclassical correction to the energy, or mass, of a kink or soliton can be computed by expanding the field around the corresponding solution,
\beq
\label{eq:Perturbation}
 \phi(t, z) = \phi_{\kink / \soliton}(z) + \vp(t,z),
\eeq
expanding the action to second order in the perturbation $\vp$ and quantizing it. The result obtained in \cite{DashenEtAl, Dashen:1975hd}, \cf also \cite{Weinberg}, for these quantities is
\begin{align}
\label{eq:E_Dashen}
  E_\kink^{\semiclass} & = \left( \frac{\sqrt{3}}{12} - \frac{3}{2 \pi} \right) m, & 
  E_\soliton^{\semiclass} & = - \frac{1}{\pi} m.
\end{align}
In order to compute the semiclassical correction, renormalization has to be performed. In \cite{DashenEtAl, Dashen:1975hd}, this is done globally: The system is first confined to a spatial interval, a formal (divergent) expression for the total energy is determined, from which a (also divergent) formal expression for the vacuum energy is subtracted. One then considers the limit in which the interval covers the whole real line, so that the mode sums become integrals (apart from a finite number of bound states). The result is still logarithmically divergent, but this divergence is cancelled by the difference of mass counterterms for the kink/soliton and the vacuum background. 

Apart from the somewhat formal nature of this renormalization procedure (in the subtraction of two divergent series, one may, by reordering, change the result by an arbitrary amount), it has the conceptual disadvantage that it is not explicitly local. But locality is a guiding principle for renormalization and crucial for its predictivity. To wit, in the absence of translation invariance (due to the presence of the kink/soliton), the local energy density can a priori be subject to arbitrary position dependent finite renormalization, unless there are clear principles restricting the renormalization freedom.

Such principles restricting the renormalization freedom were developed in the context of quantum field theory on curved spacetimes \cite{HollandsWaldWick, HollandsWaldStress, HollandsWaldReview}, where the same difficulty is present. Applying these concepts to the present case, we find that the calculation of the semiclassical corrections to the energy density is analytically feasible by elementary means, so that an infrared regularization by restriction to a finite interval is not necessary. Upon integration, we recover the above result for the semiclassical correction to the energy, if a subtlety in the definition of a conserved stress tensor for scalar fields in $1+1$ dimensions (in non-trivial backgrounds) is taken into account. Hence, we find consistency of the global approach pursued in \cite{DashenEtAl, Dashen:1975hd} and the local approach, which in our opinion is more fundamental.

The guiding principles for renormalization in curved spacetime are locality and covariance, i.e., for the renormalization of a non-linear field monomial such as $\vp^2(x)$, only the local geometric data at $x$ (or rather an infinitesimal neighborhood thereof) can be used. There are further constraints, in particular that power counting must be respected. A renormalization scheme fulfilling these requirements is \emph{Hadamard point-splitting}, where the expectation value in some state $\ket{\Omega}$ of a Wick square with derivatives is defined as
\beq
\label{eq:PointSplitRenormalization}
 \bra{\Omega} \nabla_\alpha \vp \nabla_\beta \vp(x) \ket{\Omega} \defeq \lim_{x' \to x} \nabla_\alpha \nabla'_\beta \left( \bra{\Omega} \vp(x) \vp(x') \ket{\Omega} - H(x,x') \right),
\eeq
where $\alpha$, $\beta$ are multi-indices, $\nabla'$ acts on the primed variables, and $H(x,x')$ is the \emph{Hadamard parametrix}, which is determined by the local geometric data on the geodesic connecting $x$ to $x'$. For physically reasonable states, ground states in particular, the expression in brackets on the \rhs is smooth, so that one can safely take derivatives and the limit of coinciding points. The remaining renormalization freedom is then, for each field $\nabla_\alpha \vp \nabla_\beta \vp$, finite dimensional and restricted by power counting. For example, in $3+1$ spacetime dimensions, where $\vp$ has mass dimension $1$, the finite renormalization ambiguity for $\vp^2(x)$ amounts to
\beq
 \vp^2(x) \to \vp^2(x) + \alpha R(x) + \beta \mu^2(x),
\eeq
with $R(x)$ the scalar curvature and $\mu^2(x)$ the (possibly position dependent) mass squared (the coefficient of $\vp^2$ in the Lagrangian). In particular, the Hadamard parametrix $H(x,x')$ is not unique, but contains a scale parameter in a logarithmic term, and changing that scale leads to a redefinition of this form. 

It is important to note that once the scale in the parametrix and the finite renormalization ambiguities are fixed, one has fixed a renormalization prescription on all backgrounds simultaneously, i.e., for all (globally hyperbolic) spacetimes and all possible $\mu^2(x)$. In this way, it is meaningful to say that two renormalization prescriptions on different backgrounds are the same (namely when the scale in the parametrix and all remaining finite redefinitions are the same).

Hadamard point-split renormalization is not limited to quantum field theory on curved spacetimes, but is applicable quite generally to quantum field theory in non-trivial background fields (it was in fact originally suggested by Dirac \cite{Dirac34} in the context of QED in background electromagnetic fields, see also \cite{SchlemmerZahn} for the historical context). On a flat background with no further background fields (as in the region between the plates in a Casimir setup), the Hadamard parametrix is essentially the vacuum two-point function, so the prescription \eqref{eq:PointSplitRenormalization} coincides with the usual prescription for the determination of local Casimir energy densities, see \cite{Fulling, Milton}, for example. In the context of non-trivial background fields, Hadamard point-split renormalization was for example used to compute the vacuum polarization and the energy density of fermions or scalars in a constant electric field in $1+1$ dimensions \cite{SchlemmerZahn, WernerssonZahn}, or for computing semi-classical corrections to the energy of rotating strings \cite{OpenStrings, ClosedStrings}, in agreement with results obtained previously via the Polchinski-Strominger action \cite{HellermanSwanson}. 

We note that there are other methods to compute local quantities in quantum field theory in the presence of non-trivial backgrounds \cite{GrahamEtAl, FermiPizzocchero, GoldhaberEtAl, RebhanEtAl}, but in contrast to Hadamard point-split, these have a limited range of applicability (in particular they require stationary backgrounds). In fact, the semi-classical energy density of the kink was already computed in \cite{GoldhaberEtAl} using a ``local mode regularization'' and in \cite{RebhanEtAl} via dimensional regularization. Apart from confirming this result in a renormalization scheme which is more general, our approach also provides an alternative perspective on some of the subtleties in the calculation performed in \cite{GoldhaberEtAl, RebhanEtAl}. Furthermore, we discuss in detail the treatment and the physical consequences of the zero mode, an aspect which did not receive much attention in \cite{GoldhaberEtAl, RebhanEtAl}. 

In the present setting, when quantizing the perturbations $\vp$ around the kink/soliton solution, the non-trivial background is provided by the position dependent ``mass'' 
\beq
 \mu^2_{\kink/\soliton} = V''_{\phi^4 / \sG}(\phi_{\kink / \soliton})
\eeq
occurring in the action for $\vp$. Performing an analysis according to the above principles, one finds that the only renormalization ambiguity of the Hamiltonian energy density $\rho(x)$ (defined in \eqref{eq:rho} below) is
\beq
\label{eq:rhoAmbiguity}
 \rho(x) \to \rho(x) + \alpha \mu^2(x).
\eeq
As $\mu^2_{\kink / \soliton}(z)$ converges to $m^2$ as $z \to \pm \infty$, this means that the renormalization ambiguity can be completely fixed by requiring that the renormalized semiclassical Hamiltonian energy density vanishes as $z \to \pm \infty$. Hence, with this renormalization condition, the framework yields a unique result for the semiclassical correction of the kink mass. 

The values one thus finds do, however, not coincide with \eqref{eq:E_Dashen}. This can be understood as follows. In a covariant setting, the energy density should be seen as a component of the stress tensor $T_{\mu \nu}$. The latter is classically conserved, $\del^\mu T_{\mu \nu} = 0$, but this is not guaranteed in the quantum theory in the Hadamard point-split renormalization scheme (essentially because the parametrix is in general only a solution to the equation of motion up to smooth terms). For scalar fields in spacetime dimension greater than 2,  one can achieve conservation of the renormalized stress tensor by a finite renormalization of $\del_\mu \vp \del_\nu \vp$ \cite{HollandsWaldStress}, and similarly also for Dirac fields in arbitrary dimension \cite{LocCovDirac}. However, this is not possible for the scalar field in $1+1$ dimensions. In that case, one has to slightly deviate from the principles spelled out in \cite{HollandsWaldStress, HollandsWaldReview} and allow for a finite redefinition of $T_{\mu \nu}$, independent of the Wick powers that constitute it \cite{MorettiStress}. As will be explained in detail below, performing such a redefinition, we obtain semiclassical energy densities consistent with \eqref{eq:E_Dashen}.

The article is structured as follows: In the next section, we discuss perturbations $\vp$ around the kink/soliton solutions and analyze the stress tensor at the semiclassical order. In particular, we find that the semiclassical contribution to the stress tensor splits in two terms, $T_{00}^\semiclass = \rho + T^{(1)}_{00}(\vp^{(1)})$, the first one being the Hamiltonian energy density $\rho$ of the free field in the kink/soliton background and the other one arising from taking the leading order correction to the free field into account. In Section~\ref{sec:QuantizationPerturbations}, we perform the quantization of the perturbations $\vp$ around the kink and soliton solution. In particular, we discuss in detail the effect of the ``collective coordinates'' describing position and momentum of the kink. In Section~\ref{sec:CoincidingPointDivergences} we investigate the local structure of the divergence of the Hamiltonian energy density $\rho$, i.e., the first term on the \rhs of \eqref{eq:PointSplitRenormalization}. In Section~\ref{sec:Renormalization}, we show that this divergence can be removed by a Hadamard point-split renormalization as in \eqref{eq:PointSplitRenormalization}. We discuss the remaining renormalization ambiguity and how to fix it. In Section~\ref{sec:T_1_00}, we compute the expectation value of the further contribution $T^{(1)}_{00}(\vp^{(1)})$ (corresponding to a tadpole in diagrammatic language). In Section~\ref{sec:FinalResult} we discuss the need for an additional finite renormalization of the stress tensor, in order to ensure its conservation. Combining it with the results obtained previously, we finally obtain the semiclassical contribution $T^\semiclass_{00}$ to the energy density. We conclude with a summary and an outlook.

\section{Classical solutions, their perturbations, and the stress tensor}
\label{sec:StressTensor}

The kink and the soliton solution of $\phi^4$ and sine-Gordon theory can be centered at any position $z_0$, and are given by
\begin{align}
\label{eq:phi_kink_soliton}
\phi_\kink^{z_0} & = \frac{m}{\sqrt{2 \lambda}} \tanh \frac{m (z-z_0)}{2}, &
\phi^{z_0}_\soliton & = 4 \frac{m}{\sqrt{\lambda}} \arctan ( e^{m (z-z_0)} ).
\end{align}
These solutions can also be boosted, giving rise to a kink/soliton travelling at constant velocity. The (non-boosted) solutions centered at $z_0 = 0$ will be denoted by $\phi_\kink$ and $\phi_\soliton$. 

We now consider the expansion of the Lagrangian in the perturbation $\vp$ around such a classical solution, \cf \eqref{eq:Perturbation}. To leading order in $\vp$, we find\footnote{The term linear in $\vp$ is a total derivative, as we expand around a classical solution.} (note that we use signature $(- +)$)
\beq
\label{eq:L_0}
 \cL^{(2)}_{\kink / \soliton} =  - \frac{1}{2} \del_\mu \vp \del^\mu \vp - \frac{1}{2} \mu^2_{\kink / \soliton} \vp^2.
\eeq
This will be the starting point of the semi-classical quantization that we perform in the next section. The corresponding Hamiltonian energy density is
\beq
\label{eq:rho}
 \rho_{\kink / \soliton} = \frac{1}{2} \dot \vp^2 + \frac{1}{2} \vp'^2 + \frac{1}{2} \mu^2_{\kink / \soliton} \vp^2,
\eeq
but we already anticipate that this is not the proper quantity to interpret as the energy density. Instead, we will consider the stress tensor, and for a consistent evaluation of its expectation value, we will also have to consider the next order interaction term, given by
\beq
\label{eq:L_1}
 \cL^{(3)}_{\kink / \soliton} = - \frac{1}{6} V'''_{\phi^4 / \sG}(\phi_{\kink / \soliton}) \vp^3,
\eeq
which, when evaluated for $\phi_{\kink / \soliton}$ as in \eqref{eq:phi_kink_soliton} can be seen to be of $\cO( \frac{\sqrt{\lambda}}{m})$. We will also expand the perturbation $\vp = \vp^{(0)} + \vp^{(1)} + \dots$, with the superscript denoting the order in $\frac{\sqrt{\lambda}}{m}$. In particular, $\vp^{(0)}$ is the free field subject to the equation of motion
\beq
\label{eq:eom_vp_0}
 \left( \del_\mu \del^\mu - \mu^2_{\kink / \soliton} \right) \vp^{(0)} = 0
\eeq
derived from the free part \eqref{eq:L_0} of the action, while $\vp^{(1)}$ fulfills
\beq
\label{eq:eom_vp_1}
 \left( \del_\mu \del^\mu - \mu^2_{\kink / \soliton} \right) \vp^{(1)} = \frac{1}{2} V'''_{\phi^4 / \sG}(\phi_{\kink / \soliton}) \left( \vp^{(0)} \right)^2.
\eeq

Quite generally the stress tensor for a minimally coupled\footnote{In the last section, we will briefly comment on the modifications occurring for non-minimal coupling to the scalar curvature.} scalar field $\phi$ subject to a potential $V(\phi)$ is given by
\beq
\label{eq:T_mu_nu}
 T_{\mu \nu} = \del_\mu \phi \del_\nu \phi - g_{\mu \nu} \left( \frac{1}{2} \del_\lambda \phi \del^\lambda \phi + V(\phi) \right).
\eeq
It can be obtained for example by extending the action in a minimal coupling scheme to curved spacetimes with Lorentzian metric $g_{\mu \nu}(x)$ and then varying the action \wrt $g^{\mu \nu}(x)$ (and multiplying by $\frac{-2}{\sqrt{-g}}$).
Expanding the stress tensor in the perturbation $\vp$, the leading (zeroth) order is just the classical energy density, for which one obtains
\begin{align}
\label{eq:T_00_classical}
 T_{00, \kink}^{(0)} & = \frac{1}{8} \frac{m^4}{\lambda} \sech^4 \frac{m(z-z_0)}{2}, &
 T_{00, \soliton}^{(0)} & = 4 \frac{m^4}{\lambda} \sech^2 m (z-z_0).
\end{align}
Integration over $z$ yields the classical energy of the kink/soliton as
\begin{align}
\label{eq:E_class}
 E^{\class}_\kink & = \frac{m^3}{3 \lambda}, &
 E^{\class}_\soliton & = 8 \frac{m^3}{\lambda}.
\end{align}

We now notice two important facts. First, the contribution $T^{(2)}_{00}$ of second order in the perturbation $\vp$ coincides with the Hamiltonian energy density \eqref{eq:rho}. Second, for the conservation of the stress tensor, $\del^\mu T_{\mu \nu} = 0$, it is crucial to take the full equations of motion 
\beq
 \Box \vp - \mu^2_{\kink / \soliton} \vp - \sum_{k = 3}^\infty \frac{1}{(k-1)!} V^{(k)}(\phi_{\kink / \soliton}) \vp^{k-1} = 0
\eeq
for the perturbation into account, not just the free equations of motion derived from the free action \eqref{eq:L_0}. In particular, recalling the expansion $\vp = \vp^{(0)} + \vp^{(1)} + \dots$ in the coupling $\frac{\sqrt{\lambda}}{m}$, one finds at $\cO((\frac{\sqrt{\lambda}}{m})^0)$, using \eqref{eq:eom_vp_1},
\beq
\label{eq:Divergence_T_mu_nu}
 \del^\mu \left( T_{\mu \nu}^{(1)}(\vp^{(1)}) + T_{\mu \nu}^{(2)}(\vp^{(0)}) \right) = \del_\nu \vp^{(0)} \left( \del_\mu \del^\mu - \mu^2_{\kink/\soliton} \right) \vp^{(0)}.
\eeq
The \rhs vanishes on-shell, \cf \eqref{eq:eom_vp_0}, establishing conservation of the tensor in brackets on the left hand side. We refer to it in the following as the semiclassical contribution to the stress tensor,\footnote{One can easily see that the two terms are not only of the same order in $\frac{\sqrt{\lambda}}{m}$, but also of the same order in $\hbar$: As shown below, see \eqref{eq:tilde_vp_1}, $\vp^{(1)}$ is of second order in the free field $\vp^{(0)}$, and as $T_{\mu \nu}^{(1)}(\vp^{(1)})$ is of first order in $\vp^{(1)}$, it is of second order in $\vp^{(0)}$, and hence of $\cO(\hbar)$, just like $T_{\mu \nu}^{(2)}(\vp^{(0)})$. [The tilded fields present in \eqref{eq:tilde_vp_1} are obtained from the ``untilded'' ones by omitting zero modes, a procedure explained and justified in detail below.]}
\beq
\label{eq:T_semiclass}
 T^\semiclass_{\mu \nu} = T_{\mu \nu}^{(1)}(\vp^{(1)}) + T_{\mu \nu}^{(2)}(\vp^{(0)}),
\eeq
and interpret its $\mu, \nu = 0$ component as the proper semi-classical energy density (in contrast to the Hamiltonian energy density $\rho = T_{00}^{(2)}(\vp^{(0)})$ of the free Lagrangian $\cL^{(2)}$). We will thus have to compute the expectation values of $\rho$ and
\begin{align}
 T^{(1)}_{00}(\vp^{(1)}) & = \phi'_{\kink / \soliton} \vp^{\prime (1)} + V'(\phi_{\kink / \soliton}) \vp^{(1)} \nn \\
\label{eq:T_00_1}
 & = \left( \phi'_{\kink / \soliton} \vp^{(1)} \right)'.
\end{align}
To obtain the second equality in \eqref{eq:T_00_1}, we have used that $\phi_{\kink / \soliton}$ is a time-independent solution to the classical equations of motion. In particular, this equation shows that the contribution of $T^{(1)}_{00}(\vp^{(1)})$ to the semi-classical energy density is a total derivative and does thus not contribute to the semi-classical correction of the energy. Hence, it could be ignored if our only aim was the calculation of the latter. However, as we can see from \eqref{eq:Divergence_T_mu_nu}, the inclusion of $T^{(1)}_{00}(\vp^{(1)})$ is crucial to obtain a conserved stress tensor, even at the classical level. Furthermore, for time-dependent classical background solutions (such as a breather), it has to be included to ensure the conservation of the total energy.

Finally, we note that the expectation value of the first order correction $\vp^{(1)}$ corresponds precisely to the quantum correction of the classical kink/soliton solution $\phi_{\kink/\soliton}$ first considered in \cite{ShifmanEtAl}, and denoted by $\phi_1$ there. In particular, our term $T^{(1)}_{00}(\vp^{(1)})$ corresponds to the contribution $\Delta \epsilon_{(\phi_1)}$ to the semiclassical energy density computed in \cite{GoldhaberEtAl, RebhanEtAl}.

\section{Quantization of the perturbations}
\label{sec:QuantizationPerturbations}

We now want to quantize the perturbations $\vp$, for the moment at the semi-classical level, i.e., \wrt to the free Lagrangian \eqref{eq:L_0}. In the terminology of the previous section, we are thus considering $\vp^{(0)}$. For concreteness, we first consider in detail the perturbations around the kink solution, and later state the corresponding results for the perturbations around the soliton.

In order to quantize the perturbations $\vp^{(0)}$, one considers mode solution $\vp^{(0)}(t, z) = \vp(z) e^{- i \omega t}$ to the equation of motion \eqref{eq:eom_vp_0}, i.e.,
\beq
 - \vp'' + \mu^2_\kink \vp = \omega^2 \vp.
\eeq
This has the form of a one-dimensional Schr\"odinger equation with potential $\mu^2_\kink$. As is well known (see \cite{Weinberg}, for example), this is solvable, with two bound states for $\omega_0 = 0$ and $\omega^2_1 = \frac{3}{4} m^2$ and a continuous spectrum above the asymptotic value $m^2$ of $\mu^2$, i.e., with $\omega^2_k = k^2 + m^2$ for $k \in \R$. Concretely, these are given by (we choose $z_0 = 0$ for simplicity)
\begin{align}
 \vp_0(z) & = - \frac{m^2}{\sqrt{8 \lambda}} \sech^2 \frac{m z}{2}, \\
 \vp_1(z) & = \left( \frac{3 m}{4} \right)^{\frac{1}{2}} \sech \frac{m z}{2} \tanh \frac{m z}{2}, \\
\label{eq:phi_k}
 \vp_k(z) & = \frac{1}{\sqrt{2 \pi}} \frac{ \frac{1}{2} m^2 - k^2 - \frac{3}{4} m^2 \sech^2 \frac{m z}{2} - \frac{3}{2} i k m \tanh \frac{mz}{2} }{\frac{1}{2} m^2 - k^2 - \frac{3}{2} ik m } e^{i k z}.
\end{align}
The modes $\vp_1$ and $\vp_k$ are normalized according to the usual $L^2$ norm, so in particular, $\skal{\vp_k}{\vp_{k'}} = \delta(k - k')$. The bound state $\vp_1$ corresponds to an oscillation mode of the steepness of the kink. The zero mode $\vp_0$ is due to the broken translation symmetry (and can thus be seen as a Goldstone mode) and corresponds to an infinitesimal shift of $z_0$. We have chosen its normalization such that $\vp_0 = \frac{\del}{\del z_0} \phi^{z_0}_\kink = - \frac{\del}{\del z} \phi_\kink$. It forms a symplectic pair with the linearly growing mode $t \vp_0(z)$
which arises from differentiating a boosted kink solution $\phi_\kink$ \wrt the boost rapidity (and evaluating at vanishing boost rapidity). Hence, these two modes do not behave as usual oscillator modes, but rather as position and momentum of a particle. Using these modes, the ``free'' field, i.e., the perturbation $\vp$ at lowest order, is given by
\begin{multline}
 \vp^{(0)}(t, z) = \vp_0(z) ( \hat z + \hat p t/ E_\kink^\class ) \\
 + \frac{1}{\sqrt{2 \omega_1}} \left( \vp_1(t,z) \hat a + \overline{\vp_1(t,z)} \hat a^\dag \right) + \int_{-\infty}^\infty \frac{\ud k}{\sqrt{2 \omega_k}} \left( \vp_k(t,z) \hat a_k + \overline{\vp_k(t,z)} \hat a^\dag_k \right).
\end{multline}
This fulfills the canonical equal time commutation relations iff $\hat z$ and $\hat p$ fulfill canonical commutation relations of position and momentum, while $\hat a$, $\hat a_k$ are annihilation operators, i.e.,
\begin{align}
 [\hat z, \hat p] & = i, & [ \hat a, \hat a^\dag] & = 1, & [\hat a_k, \hat a_{k'}^\dag] = \delta(k - k').
\end{align}

The operators $\hat z, \hat p$ are known as  ``collective coordinates'' representing the kink position and momentum, see \cite[Sect.~2.3]{Weinberg}, \cite[Chapter 8]{Rajaraman}, and references given there. They act as usual position and momentum operator on the Hilbert space $L^2(\R)$, representing the wave function $\psi(z)$ of the kink. As there is no normalizable ground state for these modes, it is a priori unclear how to treat them appropriately in the calculation of semi-classical corrections. To find the correct treatment (and also to justify the interpretation as kink position and momentum), we consider the interacting field $\vp_I$, which is related to the free field by 
\begin{align}
\vp_{I}(t,z)=U^{\dagger}(t)\vp^{(0)}(t,z)U(t)
\end{align}
with the evolution operator given in terms of the interaction Hamiltonian as
\begin{align}
U(t)=\hat T \exp \left[-i\int_{-\infty}^{t} H_{I}(t')\ud t'\right],
\end{align}
with $\hat T$ denoting time-ordering. 
From this expression one can extract the first order correction to the free field which leads to
\beq
\label{eq:vp_1_H_1}
\vp^{(1)}(t,z)=-i\int_{-\infty}^{t}[\vp^{(0)}(t,z), H_{1}(t')]\ud t'.
\eeq
Here $H_{1}(t)$ is the order $\frac{\sqrt{\lambda}}{m}$ part of the interaction Hamiltonian, which is given by, \cf \eqref{eq:L_1},
\beq
 H_1(t) = \frac{1}{6} \int V'''_{\phi^4}(\phi_{\kink}(z)) \left( \vp^{(0)}(t,z) \right)^3 \ud z.
\eeq
Using the fact that by the canonical equal time commutation relations, we have that
\beq
 [ \vp^{(0)}(x), \vp^{(0)}(x') ] = i (G^\ret(x,x') - G^\adv(x,x')),
\eeq
with $G^{\ret / \adv}$ retarded/advanced propagators for $\del_\mu \del^\mu - \mu^2_\kink$. Hence, as the integration in \eqref{eq:vp_1_H_1} is restricted to $t \geq t'$, we can rewrite it as
\beq
\label{eq:vp_1}
 \vp^{(1)}(t, z) = \frac{1}{2} \int G^\ret(x, x') V'''_{\phi^4}(\phi_\kink(x')) \left( \vp^{(0)}(x') \right)^2 \ud^2 x'.
\eeq
We now write
\beq
 \vp^{(0)} = - \phi'_\kink Z + \tilde \vp^{(0)},
\eeq
with
\begin{align}
 Z & = \hat z + \hat p t / E^\class_\kink, \\
\label{eq:tilde_vp_0}
 \tilde \vp^{(0)} & = \frac{1}{\sqrt{2 \omega_1}} \left( \vp_1(t,z) \hat a + \overline{\vp_1(t,z)} \hat a^\dag \right) + \int_{-\infty}^\infty \frac{\ud k}{\sqrt{2 \omega_k}} \left( \vp_k(t,z) \hat a_k + \overline{\vp_k(t,z)} \hat a^\dag_k \right),
\end{align}
i.e., $\tilde \vp^{(0)}$ is the free field up to the contributions from the zero mode. Using this expansion and\footnote{The first equation follows from $\phi_\kink$ being a static solution to the equation of motion, and the second equation is valid for all solutions to the linearized equation of motion, so in particular $\tilde \vp^{(0)}$.}
\begin{align}
 (\del_\mu \del^\mu - \mu^2_\kink) \phi_\kink'' & = V'''_{\phi^4}(\phi_\kink) (\phi'_\kink)^2, \\
 (\del_\mu \del^\mu - \mu^2_\kink) ( \tilde \vp^{(0)})' & = V'''_{\phi^4}(\phi_\kink) \phi'_\kink \tilde \vp^{(0)},
\end{align}
one finds that
\beq
 \vp^{(1)}(t, z) = \frac{1}{2} \phi''_\kink(z) Z^2 - ( \tilde \vp^{(0)})'(t, z) Z + \frac{1}{2} \int G^\ret(x, x') V'''_{\phi^4}(\phi_\kink(x')) \left( \tilde \vp^{(0)}(x') \right)^2 \ud^2 x' + \cO(\dot Z).
\eeq
The correction terms involving $\dot Z$, i.e., the momentum $\hat p$, originate from the integration by parts which is necessary to get the linearized wave operator $\del_\mu \del^\mu - \mu^2_\kink$ acting on $G^\ret$ in \eqref{eq:vp_1}. The supplementary terms omitted here implement relativistic corrections. As we will anyway be interested in states with negligible kinetic energy, these are irrelevant for our purposes. Combining our result with lower order terms, we thus have
\begin{multline}
 \phi_\kink + \vp^{(0)} + \vp^{(1)} = \left( \phi_\kink - \phi'_\kink Z + \frac{1}{2} \phi''_\kink Z^2 \right) + \left( \tilde \vp^{(0)} - (\tilde \vp^{(0)})' Z \right) \\ + \frac{1}{2} \int G^\ret(x, x') V'''_{\phi^4}(\phi_\kink(x')) \left( \tilde \vp^{(0)}(x') \right)^2 \ud^2 x' + \cO(\dot Z).
\end{multline}
This looks like a Taylor expansion in $Z$, and indeed one can show \cite{MatsumotoEtAl1979} that upon including all orders in $Z$ one obtains
\beq
\label{eq:phi_Q_Expansion}
 \phi(t,z) = \phi_\kink(t, z - Z) + \tilde \vp^{(0)}(t, z - Z) + \tilde \vp^{(1)}(t, z - Z) + \cO(\dot Z, \sqrt{\lambda})
\eeq
with
\beq
\label{eq:tilde_vp_1}
 \tilde \vp^{(1)}(t, z) = \frac{1}{2} \int G^\ret(x, x') V'''_{\phi^4}(\phi_\kink(x')) \left( \tilde \vp^{(0)}(x') \right)^2 \ud^2 x'.
\eeq
We thus see that $\hat z$, $\hat p$ indeed play the role of collective coordinates describing the position and momentum of the kink, as they implement a shift in the position and momentum of the classical kink, as well as of the perturbations around it. The contributions of order $\dot Z$ implement relativistic corrections. Such a connection between the perturbative and the relativistic expansion has already been noted in \cite{Gervais:1975pa}, where it was also argued that including $\dot Z$ to all orders precisely implements relativistic dispersion relations, at least as far as the total energy is concerned.

Let us now consider the contribution of the first term on the \rhs of \eqref{eq:phi_Q_Expansion} to the expectation value of the stress tensor component $T_{00}$. As it only depends on the collective coordinates $\hat z$, $\hat p$, we only need to evaluate it in a state $\psi \in L^2(\R)$, i.e., a wave function for the position of the kink. We obtain
\beq
\label{eq:T_00_CollCoords}
 \bra{\psi} T_{00, \mathrm{Coll.Coor.}}(z, t) \ket{\psi} = \bra{\psi} T^{(0)}_{00, \kink}(z - Z) \ket{\psi} + \frac{1}{2 (E^\class_\kink)^2} \bra{\psi} ( \{ \hat p, \phi'_\kink(z-Z) \}_{\mathrm{sym}} )^2 \ket{\psi}.
\eeq
In the first term on the r.h.s., we have the classical contribution \eqref{eq:T_00_classical} to the energy density, with $z-z_0$ replaced by $z-Z$. In the second term on the r.h.s., an operator ordering symmetrization has to be performed (for example by expanding $\phi'_\kink$ as a power series and in each term fully symmetrize $\hat p$ with all $Z$'s). 

If for simplicity we restrict to $t = 0$, and recalling that $\hat z$ acts as multiplication operator on the wave function $\psi(z)$, we see that the first term on the \rhs of \eqref{eq:T_00_CollCoords} is simply the convolution of the classical energy density with the probability density $| \psi (z) |^2$. This convolution does obviously not affect the total energy. At other times $t$, one convolutes with $| \psi_t(z) |^2$ instead, where $\psi_t$ is obtained from $\psi$ by acting with the time evolution operator corresponding to the free Hamiltonian $\frac{1}{2 E^\class_\kink} \hat p^2$.

As for the second term on the \rhs of \eqref{eq:T_00_CollCoords}, it contains a factor $\hat p^2$, so this term is a contribution due to the kinetic energy of the kink. Choosing a wave function $\psi(z)$ which is sufficiently broad, this contribution can be made arbitrarily small, and should thus be neglected in the evaluation of the energy density in a ground state.\footnote{The consideration of a ``broad'' wave function can also be motivated as follows: Introducing a spatial cut-off, the zero mode turns into a mode with a small finite frequency (which converges to 0 as the cut-off is moved to infinity). Mathematically, the amplitude of such a mode is described by a harmonic oscillator and the ground state in a Schr\"odinger representation (for the mode amplitude) is a Gaussian wave function. In the limit where the spatial cut-off is moved to infinity, the width of the Gaussian diverges (the limit thus does not define a normalizable wave function). The wave functions $\psi(z)$ that we are considering are precisely such ``broad'' wave functions for which the variance of $\hat p$ (and thus the kinetic contribution to the energy density) can be neglected.} For general wave functions $\psi(z)$ the contribution of this term to the expectation value of the total energy is $\frac{1}{2 E^\class_\kink} \bra{\psi} \hat p^2 \ket{\psi}$, as expected for the kinetic energy. Analogous considerations apply to the other contributions to the energy density, i.e., those involving also the second and third term on the \rhs of \eqref{eq:phi_Q_Expansion}.

Hence, up to a contribution from the kinetic energy, which can be made arbitrarily small, the effect of the collective coordinates is to smear (convolute) the classical energy density and its semiclassical correction (computed with the ``tilded'' fields) with the probability density $| \psi (z) |^2$ for the position of the kink. As there is no preferred choice (ground state) for the wave function $\psi(z)$, and we do not intend to introduce an ad-hoc choice, this is as much as can be said about the influence of the collective coordinates. We will thus continue with the evaluation of the semi-classical corrections due to the ``tilded'' perturbations, i.e., $T_{00}^{(0)}(\tilde \vp^{(0)})$ and $T_{00}^{(1)}(\tilde \vp^{(1)})$. The basic ingredient for the computation of their contribution to the semi-classical energy density is the two-point function of $\tilde \vp^{(0)}$ in the ground state $\ket{\Omega_\kink}$ which is defined to be the state annihilated by $\hat a$, $\hat a_k$:
\beq
 \bra{\Omega_\kink} \tilde \vp^{(0)}(x) \tilde \vp^{(0)}(x') \ket{\Omega_\kink} = \frac{1}{2 \omega_1} \vp_1(z) \overline{\vp_1(z')} e^{- i \omega_1(t-t')} + \int_{-\infty}^\infty \frac{\ud k}{2 \omega_k} \vp_k(z) \overline{\vp_k(z')} e^{- i \omega_k (t-t')}.
\eeq

\section{Evaluating $T^{(2)}_{00}(\tilde \vp^{(0)})$}
\label{sec:CoincidingPointDivergences}

As discussed above, \cf \eqref{eq:T_semiclass}, there are two contributions to the semiclassical energy density, the Hamiltonian energy density $T_{00}^{(2)}(\tilde \vp^{(0)})  = \rho$ corresponding to the free action, and $T_{00}^{(1)}(\tilde \vp^{(1)})$. We now focus on the first (the second one will be treated in Section~\ref{sec:T_1_00}). 
A naive evaluation of the expectation value of $T_{00}^{(2)}(\tilde \vp^{(0)})$ will be divergent. In order to prepare for the Hadamard point-split renormalization to be performed in the next section, we regularize it by a point-split, i.e., in the products of fields occurring in \eqref{eq:rho}, we evaluate the first field at $x = (t, z)$ and the second one at $x' = (t + \tau, z)$. Such a point-split in the time direction is typically advantageous in static backgrounds. The relevant expression is thus
\beq
 \rho^{\ps}(z; \tau) = \frac{1}{2} \left( \del_{t} \del_{t'} + \del_z \del_{z'} + \mu^2(z) \right) \bra{\Omega} \tilde \vp^{(0)}(x) \tilde \vp^{(0)}(x') \ket{\Omega}.
\eeq
A straightforward calculation gives\footnote{These and all the other calculations relevant for the kink background are performed in a \textsc{Mathematica} notebook that is associated to the arXiv submission as an ancillary file.} (we introduced an $i \eps$ prescription to ensure convergence)
\begin{align}
\label{eq:rho_kink_Integral}
 \rho^{\ps}_\kink(z; \tau) & = \frac{\sqrt{3}}{32} m^2 \left( 10 \sech^6 \frac{mz}{2} - 17 \sech^4 \frac{mz}{2} + 8 \sech^2 \frac{mz}{2} \right) e^{i \omega_1 \tau } \\
 & \quad + \int_{-\infty}^\infty \frac{\ud k}{8 \pi} \frac{e^{i \omega_k (\tau + i \eps)}}{\omega_k^3 (4 k^2 + m^2)} \bigg[ 2 \omega_k^4 (4 k^2 + m^2) - \frac{3}{2} m^2 \omega_k^2 \left( 4 k^2 + 5 m^2 \right) \sech^2 \frac{mz}{2} \nn \\
 & \qquad \qquad  + \frac{9}{4} m^4 \left( 3 k^2 + 5 m^2 \right) \sech^4 \frac{m z}{2} - \frac{45}{8} m^6 \sech^6 \frac{m z}{2} \bigg] . \nn
\end{align}
The first summand comes from the $\omega_1$ mode. In that term, the coinciding point limit $\tau \to 0$ can be taken. Integration of this term over $z$ then gives the contribution $\frac{1}{2} \omega_1$ to the semiclassical energy, consistent with the expectation for a single oscillator mode. To treat the integral term in \eqref{eq:rho_kink_Integral}, we first perform a change of variables to write it as
\beq
\label{eq:int_I}
 \int_{m}^\infty I(\omega, z) e^{i \omega (\tau + i \eps)} \ud \omega,
\eeq
with $I(\omega, z)$ a polynomial in $\sech^2 \frac{m z}{2}$ with functions of $\omega$ as coefficients. We perform an asymptotic expansion of $I(\omega, z)$ for large $\omega$, yielding
\beq
 I(\omega, z) = \frac{1}{2 \pi} \omega + \frac{\mu^2_\kink(z)}{4 \pi} \frac{1}{\omega} + R(\omega, z),
\eeq
where the remainder $R(\omega, z)$ is of $\cO(\omega^{-3})$. We also used that 
\beq
\label{eq:mu2_kink}
 \mu_\kink^2(z) = m^2 \left( 1 - \frac{3}{2} \sech^2 \frac{m z}{2} \right).
\eeq
We now write the integral \eqref{eq:int_I} as
\beq
 \int_{m}^\infty R(\omega, z) e^{i \omega (\tau + i \eps)} \ud \omega + \int_{m}^\infty \left( \frac{1}{2 \pi} \omega + \frac{\mu_\kink^2(z)}{4 \pi} \frac{1}{\omega} \right) e^{i \omega (\tau + i \eps)} \ud \omega.
\eeq
The integrand of the first term is absolutely convergent, so we may take the coinciding point limit $\tau \to 0$ directly inside the integral, which can then be performed analytically. To deal with the second term, we note that
\begin{align}
 \int_{m}^\infty \omega e^{i \omega (\tau + i \eps)} \ud \omega & = \frac{-1}{(\tau + i \eps)^2} - \frac{m^2}{2} + \cO(\tau), \\
 \int_{m}^\infty \frac{1}{\omega} e^{i \omega (\tau + i \eps)} \ud \omega & = - \frac{1}{2} \log [ - m^2 (\tau + i \eps)^2 ] - \gamma + \cO(\tau),
\end{align}
with $\gamma$ the Euler-Mascheroni constant. In this way, we have completely determined the divergence in the coinciding point limit, including the finite term. Combining all the contributions, we find
\begin{align}
 \rho^\ps_\kink(z; \tau) & = - \frac{1}{2 \pi ( \tau + i \eps )^2} - \frac{1}{8 \pi} \mu_\kink^2(z) \left(  \log [ - m^2 (\tau + i \eps)^2 ] + 2 \gamma + 1 - \log 4 \right)  \nn \\
 & \quad + \left( \frac{\sqrt{3}}{12} - \frac{3}{16 \pi} \right) m^2 \sech^2 \frac{mz}{2} - \left( \frac{17 \sqrt{3}}{96} + \frac{3}{8 \pi} \right) m^2 \sech^4 \frac{m z}{2} \nn \\
\label{eq:rho_ps}
 & \quad + \left( \frac{5 \sqrt{3}}{48} + \frac{15}{32 \pi} \right) m^2 \sech^6 \frac{m z}{2} + \cO(\tau).
\end{align}
Here, we have written the terms which are finite in the coinciding point limit $\tau \to 0$ and vanish as $z \to \pm \infty$ in the last two lines. In the first line, we find a leading quadratic divergence, which is $z$ independent, and a logarithmic divergence, which is proportional to the position-dependent ``mass'' $\mu_\kink^2(z)$. As we will see, this is no coincidence, but reflects the Hadamard property of the ground state, i.e., that the singularities of the two-point function are captured by the Hadamard parametrix.

For the evaluation of the second contribution to the semiclassical energy density, $T_{00}^{(1)}(\tilde \vp^{(1)})$, we will need also the (renormalized) expectation value of $(\tilde \vp^{(0)})^2(x)$. In a point-split prescription, this amounts to computing the two-point function, i.e., we consider
\beq
 \vp^{2, \ps}(z; \tau) = \bra{\Omega} \tilde \vp^{(0)}(x) \tilde \vp^{(0)}(x') \ket{\Omega},
\eeq
where again $x= (t, z)$ and $x' = (t + \tau, z)$. Proceeding as before, one obtains
\begin{multline}
\label{eq:p2_ps}
 \vp^{2, \ps}_\kink(z; \tau) = - \frac{1}{4 \pi} \left( \log [ - m^2 (\tau + i \eps)^2 ] + 2 \gamma - \log 4 \right) \\
 + \frac{\sqrt{3}}{12} \sech^2 \frac{mz}{2} - \left( \frac{3}{8 \pi} + \frac{\sqrt{3}}{12} \right) \sech^4 \frac{mz}{2}  + \cO(\tau).
\end{multline}

Let us now briefly state the corresponding expressions for the perturbations around the soliton. 
We note that
\beq
\label{eq:mu2_soliton}
 \mu^2_\soliton = m^2 \left( 1 - 2 \sech^2 m z \right),
\eeq
so the differential equation for the modes is very similar to the one for the kink perturbations, for which we had \eqref{eq:mu2_kink}. In contrast to the kink, there is now only the zero mode (due to translational symmetry) as a bound state. The normalized continuum modes are given by
\beq
 \vp_k(z) = \frac{1}{\sqrt{2 \pi}} \frac{i k - m \tanh mz}{i k - m} e^{i k z},
\eeq
again with frequency $\omega_k = \sqrt{k^2 + m^2}$. As for the perturbations around the kink, the zero mode gives rise to collective coordinates, which are treated in complete analogy to the above. Following the same steps as for the perturbations around the kink, one obtains
\begin{align}
\label{eq:rho_soliton_ps}
 \rho^\ps_\soliton & = \frac{-1}{2 \pi (\tau + i \eps)^2} - \frac{1}{8 \pi} \mu^2_\soliton(z) \left(  \log [ - m^2 (\tau + i \eps)^2 ] + 2 \gamma + 1 - \log 4 \right) \nn \\
 & \quad - \frac{3}{4 \pi} m^2 \sech^2 m z + \frac{3}{4 \pi} m^2 \sech^4 m z + \cO(\tau), \\
\label{eq:p2_soliton_ps}
 \vp^{2, \ps}_\soliton & = - \frac{1}{4 \pi} \left( \log [ - m^2 (\tau + i \eps)^2 ] + 2 \gamma - \log 4 \right) - \frac{1}{2 \pi} \sech^2 m z + \cO(\tau).
\end{align}

\section{Hadamard point-split renormalization}
\label{sec:Renormalization}

According to the Hadamard point-split description \eqref{eq:PointSplitRenormalization}, we have to subtract from the point-split Hamiltonian energy densities just computed the corresponding expression derived from the Hadamard parametrix.
In $1+1$ dimensions, the Hadamard parametrix is of the form (for an overview, see \cite{DecaniniFolacci}, for example)
\beq
\label{eq:H}
 H(x,x') = - \frac{1}{4 \pi} V(x,x') \log \frac{ \sigma + i \eps (t - t') }{\Lambda^2}
\eeq
with $\Lambda$ a length scale, whose choice will amount to a choice of renormalization condition, $\sigma(x,x')$ being $\frac{1}{2}$ times the squared geodesic distance of $x$ and $x'$, i.e., on Minkowski space, $\sigma = \frac{1}{2} ( (z-z')^2 - (t-t')^2 )$ and $V(x,x')$ a smooth function which can be expanded in terms of Hadamard coefficients $V_k(x,x')$ (Lorentzian analogs of heat kernel coefficients) as
\beq
 V(x, x') = \sum_{k = 0}^\infty V_k(x,x') \sigma^k(x,x').
\eeq
The Hadamard coefficients are determined uniquely by the geometric data occurring in the wave equation governing the dynamics of the field. $V_0$ is the square root of the vanVleck determinant, i.e., $V_0 = 1$ on Minkowski space. For our purposes, i.e., the evaluation of \eqref{eq:PointSplitRenormalization}, we further only need the coinciding point limit of $V_1$, which is $V_1(x,x) = \frac{1}{2} \mu^2(x)$ (all other contributions are at least of third order in $x'-x$ and are thus not relevant for the coinciding point limit performed in \eqref{eq:PointSplitRenormalization}). Performing the necessary derivatives and again evaluating at a point-split in time direction, $x' = (t + \tau, z)$, one straightforwardly finds
\begin{align}
\label{eq:rho_H}
  \frac{1}{2} \left( \del_{t} \del_{t'} + \del_z \del_{z'} + \mu^2(z) \right) H(x, x') & = - \frac{1}{2 \pi ( \tau + i \eps )^2} - \frac{1}{8 \pi} \mu^2(z) \left(  \log \frac{- (\tau + i \eps)^2}{2 \Lambda^2} + 1 \right) + \cO(\tau) , \\
\label{eq:p2_H}
  H(x, x') & = - \frac{1}{4 \pi} \log \frac{- (\tau + i \eps)^2}{2 \Lambda^2} + \cO(\tau).
\end{align}
We see that the divergences of \eqref{eq:rho_ps} and \eqref{eq:p2_ps} (and also \eqref{eq:rho_soliton_ps}, \eqref{eq:p2_soliton_ps} in the soliton background) in the coinciding point limit are cancelled by this subtraction. One thus obtains locally finite expectation values of the renormalized Hamiltonian energy density $\rho$ and $(\tilde \vp^{(0)})^2$. Furthermore, we can fix the scale $\Lambda$ by requiring that the Hamiltonian energy density $\rho$ vanishes as $z \to \pm \infty$, i.e., by setting $\Lambda = \sqrt{2}(e^{\gamma} m)^{-1}$. Obviously, with the same value, also the renormalized expectation value of $(\tilde \vp^{(0)})^2(z)$ vanishes as $z \to \pm \infty$. In particular, for the expansion around a classical vacuum solution, one finds, for this value of $\Lambda$, vanishing expectation values both of the Hamiltonian energy density $\rho$ and of $(\tilde \vp^{(0)})^2$. In this sense, in the vacuum background, our renormalization condition effectively coincides with normal ordering of the free Hamiltonian.

As discussed in the Introduction, even after a Hadamard point-split renormalization, there is in principle some further, but highly constrained renormalization freedom. In $1+1$ dimensions, and in the absence of curvature, it consists in redefinitions
\begin{align}
\label{eq:redef_p2}
 \vp^2 & \to \vp^2 + \alpha, \\
\label{eq:redef_del_p2}
 \del_\mu \vp \del_\nu \vp & \to \del_\mu \vp \del_\nu \vp + \beta g_{\mu \nu} \mu^2,
\end{align}
with $\alpha, \beta \in \R$, as a constant and the ``mass'' squared are the only local geometric quantities (occurring in the free action) of mass dimension $0$ and $2$, respectively. While \eqref{eq:redef_del_p2} does not affect the Hamiltonian energy density $\rho$, \cf \eqref{eq:rho} [and obviously also the not Wick square $\vp^2$], the redefinition \eqref{eq:redef_p2} amounts to adding $\frac{\alpha}{2} \mu^2(z)$ to the Hamiltonian energy density $\rho$. But this is tantamount to the same modification of $\Lambda$ in the subtraction terms \eqref{eq:rho_H} and \eqref{eq:p2_H}. Hence, the requirement of vanishing energy density at $z \to \pm \infty$ fixes completely the renormalization ambiguity for that quantity. Subtracting \eqref{eq:rho_H} and \eqref{eq:p2_H} from \eqref{eq:rho_ps} and \eqref{eq:p2_ps}, we obtain for the expectation values of the renormalized $\rho$ and $(\tilde \vp^{(0)})^2$ in the kink background
\begin{align}
 \rho^\semiclass_\kink(z) & = \left( \frac{\sqrt{3}}{12} - \frac{3}{16 \pi} \right) m^2 \sech^2 \frac{mz}{2} - \left( \frac{17 \sqrt{3}}{96} + \frac{3}{8 \pi} \right) m^2 \sech^4 \frac{m z}{2} \nn \\
\label{eq:rho_semiclass}
 & \quad + \left( \frac{5 \sqrt{3}}{48} + \frac{15}{32 \pi} \right) m^2 \sech^6 \frac{m z}{2}, \\
\label{eq:p2_semiclass}
 \vp^{2, \semiclass}_\kink(z) & = \frac{\sqrt{3}}{12} \sech^2 \frac{mz}{2} - \left( \frac{3}{8 \pi} + \frac{\sqrt{3}}{12} \right) \sech^4 \frac{mz}{2}. 
\end{align}
Similarly, in the soliton background, one obtains, from \eqref{eq:rho_soliton_ps}, \eqref{eq:p2_soliton_ps},
\begin{align}
\label{eq:rho_soliton_semiclass}
 \rho^\semiclass_\soliton(z) & = - \frac{3}{4 \pi} m^2 \sech^2 mz + \frac{3}{4 \pi} m^2 \sech^4 m z, \\
\label{eq:p2_soliton_semiclass}
 \vp^{2, \semiclass}_\soliton(z) & = - \frac{1}{2 \pi} \sech^2 m z. 
\end{align}

\section{Evaluating $T^{(1)}_{00}(\tilde \vp^{(1)})$}
\label{sec:T_1_00}

While in the previous sections, we have calculated the Hamiltonian energy density $\rho = T^{(2)}_{00}(\tilde \vp^{(0)})$, we now turn to the evaluation of $T^{(1)}_{00}(\tilde \vp^{(1)})$, which, as seen in \eqref{eq:T_00_1} is linear in $\tilde \vp^{(1)}$. We recall the explicit expression \eqref{eq:tilde_vp_1} for the latter. For its expectation value, we thus obtain
\beq
\label{eq:EV_vp_1}
\bra{\Omega_{\kink}} \tilde \vp^{(1)}(t,z) \ket{\Omega_{\kink}} = \frac{1}{2} \int G^\ret(x, x') V_{\phi^4}'''(\phi_{\kink}(z')) \vp^{2, \semiclass}_\kink(z') \ud^2 x',
\eeq
where we substituted the renormalized value for the expectation value $\vp^{2, \semiclass}_\kink(z')$, which is time independent. The retarded propagator $G^\ret$ can be expressed in terms of the modes as
\begin{multline}
 G^\ret(x, x') = \Theta(t- t') \Bigg[ \vp_0(z) \vp_0(z') \frac{t' - t}{E^\class} \\
 + \Imag \left( \frac{1}{\omega_1} \vp_1(z) \overline{\vp_1(z')} e^{- i \omega_1 (t-t')} + \int \ud k \frac{1}{\omega_k} \vp_k(z) \overline{\vp_k(z')} e^{- i \omega_k (t-t')} \right) \Bigg],
\end{multline}
with $\Theta$ the step function. As both $\vp_0(z')$ and $ \vp^{2, \semiclass}_\kink(z')$ are even functions of $z'$, while $V_{\phi^4}'''(\phi_{\kink}(z'))$ is odd, the first term on the \rhs does not contribute to \eqref{eq:EV_vp_1}. In the remaining terms, the integration over $t'$ can be performed (with the imposition of a suitable $i \eps$ prescription). An analogous expression (without the term involving $\vp_1$) holds in the soliton background. In both cases the remaining integrals (over $k$ and $z'$) can be performed analytically, and one obtains, for the expectation value of $\tilde \vp^{(1)}$,
\begin{align}
 \bra{\Omega_\kink}\tilde \vp^{(1)}(z) \ket{\Omega_\kink} & = \frac{\sqrt{2\lambda}}{2m}\left[\left(\frac{\sqrt{3}}{6}+\frac{3}{8\pi}\right)\tanh\frac{mz}{2}-\frac{\sqrt{3}}{8}mz\right]\sech^2\frac{mz}{2}, \\
\label{eq:phi_1_soliton}
 \bra{\Omega_\soliton}\tilde \vp^{(1)}(z) \ket{\Omega_\soliton} & = \frac{\sqrt{\lambda}}{8\pi m}\tanh mz\sech mz.
\end{align}
With these expressions one obtains for the expectation value of $T^{(1)}_{00}(\tilde \vp^{(1)})(z)$ at the relevant order, by using \eqref{eq:T_00_1},
\begin{align}
\label{eq:T_1_kink}
 \bra{\Omega_\kink} T^{(1)}_{00}(\tilde \vp^{(1)})(z) \ket{\Omega_\kink} & = -\left(\frac{11\sqrt{3}}{96}+\frac{3}{8\pi}\right) m^2 \sech^4 \frac{m z}{2} + \left(\frac{5\sqrt{3}}{48}+\frac{15}{32\pi}\right) m^2 \sech^6 \frac{m z}{2} \nn \\ 
 & \quad + \frac{\sqrt{3}}{16} m^3 z \tanh \frac{m z}{2} \sech^4 \frac{m z}{2}, \\
\label{eq:T_1_soliton}
 \bra{\Omega_\soliton} T^{(1)}_{00}(\tilde \vp^{(1)})(z) \ket{\Omega_\soliton} & = - \frac{1}{2 \pi} m^2 \sech^2 m z + \frac{3}{4 \pi} m^2 \sech^4 m z.
\end{align}
Our result for the kink coincides with the contribution $\Delta \epsilon_{(\phi_1)}$ to the semiclassical energy density computed in \cite{GoldhaberEtAl, RebhanEtAl}. 
Furthermore, our result \eqref{eq:phi_1_soliton} for the expectation value of $\tilde \vp^{(1)}$ for the soliton coincides with the result for $\phi_1$ computed in \cite{RebhanEtAl} (the semiclassical energy density of the soliton was not computed there).

\section{The semi-classical energy densities}
\label{sec:FinalResult}

As discussed in Section~\ref{sec:StressTensor}, the proper quantity to describe semi-classical corrections to the energy density is $T^\semiclass_{00} = T_{00}^{(1)}(\tilde \vp^{(1)}) + T_{00}^{(2)}(\tilde \vp^{(0)})$, and we have computed the renormalized expectation values of both terms in the previous sections. However, it turns out that the renormalized expectation value of the \rhs of \eqref{eq:Divergence_T_mu_nu} does not vanish in the situation we are considering. Hence, we still need to introduce a correction term in order to ensure a vanishing divergence of the renormalized stress tensor.

For the \rhs of \eqref{eq:Divergence_T_mu_nu} in the quantum theory, using Hadamard point split renormalization, one finds \cite{MorettiStress} (we do not have to take an expectation value, as this is a c-number)
\beq
 \del_\nu \tilde \vp^{(0)} ( \del_\mu \del^\mu - \mu^2 ) \tilde \vp^{(0)} = - \lim_{x' \to x} ( \del_\mu \del^\mu - \mu^2 ) \del'_\nu H(x, x') = \frac{1}{4 \pi} \del_\nu V_1(x, x),
\eeq
where on the \rhs the derivative acts on both variables. As $V_1(x,x) = \frac{1}{2} \mu^2(x)$ is not constant in the cases discussed here, our condition for the conservation of energy is violated.
From \eqref{eq:eom_vp_1}, one straightforwardly finds that
\beq
 \del^\mu T^{(1)}_{\mu \nu} (\tilde \vp^{(1)}) = \frac{1}{2} \del_\nu \mu^2 \left( \tilde \vp^{(0)} \right)^2.
\eeq
It follows that the redefinitions \eqref{eq:redef_p2}, \eqref{eq:redef_del_p2} of $(\tilde \vp^{(0)})^2$ and $\del_\mu \tilde \vp^{(0)} \del_\nu \tilde \vp^{(0)}$ leave the \lhs of \eqref{eq:Divergence_T_mu_nu} invariant, so these can not be used to achieve a conserved stress tensor. This is a special property of scalar field theory in $1+1$ dimensions, not present in higher dimensions \cite{HollandsWaldStress} or for Dirac fields \cite{LocCovDirac}. A way out, proposed in \cite{MorettiStress}, is to directly modify the stress tensor as $T^{(2)}_{\mu \nu}(x) \to T^{(2)}_{\mu \nu}(x) - \frac{1}{4 \pi} g_{\mu \nu} V_1(x,x)$. As $\mu^2(z)$ does not vanish as $z \to \infty$, we modify this slightly by subtracting the asymptotic value at $\infty$, i.e., we instead consider
\beq
\label{eq:T_mu_nu_redefinition}
 T^{(2)}_{\mu \nu}(x) \to T^{(2)}_{\mu \nu}(x) - \frac{1}{8 \pi} g_{\mu \nu} \left( \mu^2(z) - m^2 \right).
\eeq
This can also be interpreted as considering the difference of the modified $T$'s in the kink/soliton and the vacuum background. A generalization of this modification to the non-perturbative sine-Gordon theory was independently proposed in \cite{CadamuroFrob}.

\begin{figure}
\centering
\includegraphics{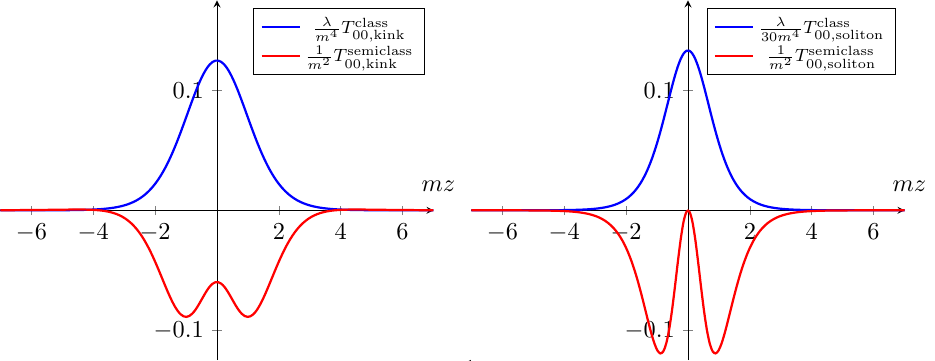}
\caption{Comparison of the classical energy densities \eqref{eq:T_00_classical} [in blue] and the semiclassical corrections \eqref{eq:T00_kink}, \eqref{eq:T00_soliton} [in red]. Note that both have been scaled by the appropriate powers of $\lambda$ and $m^{-2}$. Furthermore, $T^\class_{00,\soliton}$ has been further scaled by a factor $\frac{1}{30}$ for better readability.}
\label{fig:rho}
\end{figure}

Combining \eqref{eq:rho_semiclass}, \eqref{eq:rho_soliton_semiclass} with \eqref{eq:T_1_kink}, \eqref{eq:T_1_soliton} and the redefinition \eqref{eq:T_mu_nu_redefinition}, we finally obtain the semiclassical energy densities
\begin{align}
 T^{\semiclass}_{00,\kink}(z) & =\left( \frac{\sqrt{3}}{12} - \frac{3}{8 \pi} \right) m^2 \sech^2 \frac{mz}{2} - \left( \frac{28 \sqrt{3}}{96} + \frac{3}{4 \pi} \right) m^2 \sech^4 \frac{m z}{2} \nn \\
\label{eq:T00_kink}
 & \quad + \left( \frac{5 \sqrt{3}}{24} + \frac{15}{16 \pi} \right) m^2 \sech^6 \frac{m z}{2} + \frac{\sqrt{3}}{16} m^3 z \tanh \frac{m z}{2} \sech^4 \frac{m z}{2}, \\
\label{eq:T00_soliton}
 T^{\semiclass}_{00,\soliton}(z) & = - \frac{3}{2\pi} m^2 \sech^2 mz + \frac{3}{2 \pi} m^2 \sech^4 m z. 
\end{align}
These are plotted in Fig.~\ref{fig:rho}, together with the corresponding classical energy densities. Upon integration of \eqref{eq:T00_kink}, \eqref{eq:T00_soliton} over $z$, one finds the semiclassical correction \eqref{eq:E_Dashen} to the energy computed in \cite{DashenEtAl, Dashen:1975hd}. However, had we omitted the correction term \eqref{eq:T_mu_nu_redefinition} to ensure conservation of the stress tensor, we would have obtained different results.

It is noteworthy that the first term in \eqref{eq:T00_kink} dominates at sufficiently large distances over the classical contribution \eqref{eq:T_00_classical}, due to the lower power of $\sech \frac{m z}{2}$. In view of quantum energy inequalities \cite{Fewster:1998pu} forbidding negative energy densities for arbitrarily large times, it is thus reassuring that the coefficient $\frac{\sqrt{3}}{12} - \frac{3}{8 \pi} \simeq 0.025$ is positive (albeit quite small, so that the longer range of the semi-classical correction is not perceptible in Fig.~\ref{fig:rho}).

Our result \eqref{eq:T00_kink} for the kink was previously obtained in \cite{GoldhaberEtAl}, using ``local mode regularization''. In that procedure, a correction term, termed ``anomaly'' is involved. In a dimensional regularization approach \cite{RebhanEtAl}, it occurs as an $\epsilon / \epsilon$ term (and would be missing in a naive momentum cutoff regularization). This ``anomaly'' term precisely corresponds to our redefinition \eqref{eq:T_mu_nu_redefinition}, see also the concluding discussion in \cite{GoldhaberEtAl}. With that correction, our result for $T^{(2)}_{00}(\tilde \vp^{(0)})$ for the kink coincides with the ``local Casimir density'' $\epsilon_{\mathrm{Cas}} + \Delta \epsilon_{\mathrm{Cas}}$ computed in \cite{GoldhaberEtAl}.

We are not aware of published results for the semiclassical energy density of the soliton in the sine-Gordon model. However, in \cite{ShifmanEtAl} a result for the supersymmetric sine-Gordon model is given, which is qualitatively of the same form as \eqref{eq:T00_soliton}, i.e., a linear combination of $\sech^2 m z$ and $\sech^4 m z$. Furthermore, as already mentioned, the expectation value of $\tilde \vp^{(1)}$ in the sine-Gordon model was computed in \cite{RebhanEtAl}, consistently with our result \eqref{eq:phi_1_soliton}.

We remark that in \cite{Mukhopadhyay:2021wmu} the energy density in a similar model was computed numerically. There, a massive free scalar field coupled to a background sine-Gordon field was considered. The latter was taken to be a kink-antikink scattering solution, but at early times the two kinks can be considered as isolated, and thus the situation is comparable to ours (the Lagrangian considered is essentially \eqref{eq:L_0} with an additional mass term). In that situation, the contribution $T^{(1)}_{00}(\tilde \vp^{(1)})$ is absent, so only the Hamiltonian energy density needs to be considered. The plots shown in \cite{Mukhopadhyay:2021wmu} seem to qualitatively differ from our result for the latter, but whether this is due to the presence of a supplementary mass term or due to a different renormalization scheme (the modification \eqref{eq:T_mu_nu_redefinition}, which was crucial for consistency with the global results, was apparently not implemented) is presently unclear.

We also note that in \cite{Baacke}, the similar (but more considerably demanding) problem of computing semiclassical energy densities for the Nielsen-Olesen vortex was tackled. However, apparently the requirement of conservation of the stress tensor was not implemented, as neither the contribution $T^{(1)}_{00}(\tilde \vp^{(1)})$ needed at the classical level, nor the finite renormalization necessary to ensure conservation of the stress tensor at the quantum level seem to have been taken into account.

\section{Discussion \& Outlook}

Several remarks on our results and perspectives on future research are in order:

\begin{itemize}

\item Situations in which the renormalized energy density can be computed analytically in non-trivial backgrounds are quite rare, so our results provide useful examples for the application of the local and covariant renormalization technique in physically relevant settings.

\item We used conventions such that $m$ is the mass corresponding to classical fluctuations around the vacuum $\phi_\vac$. In the quantum theory, the mass of the associated particle is subject to self-energy corrections. With the renormalization conditions used here (where the tadpole vanishes in the vacuum background), only the $\phi^3$ ``setting sun'' graph contributes to the self-energy at the one-loop level, and one obtains, for the $\phi^4$ model, the physical mass $\bar m = m ( 1 - \frac{\sqrt{3}}{4} \frac{\lambda}{m^2} + \cO(\frac{\lambda}{m^2}) )$ \cite{Rebhan:1997iv} (the sine-Gordon model does not have a $\phi^3$ term, so it is not subject to such a modification). Using a correspondingly adjusted coupling constant $\bar \lambda = \frac{\bar m^2}{m^2} \lambda$ it is natural to express the results for the energy density in terms of these quantities, see \cite{GoldhaberEtAl}.

\item By introducing a non-minimal coupling $-\xi R \phi^2$ to the scalar curvature in the Lagrangian, the stress tensor is modified as
\begin{equation}
 T_{\mu \nu} \to T_{\mu \nu} + \xi \left( \eta_{\mu \nu} \del^\lambda \del_\lambda - \del_\mu \del_\nu \right) \phi^2
\end{equation}
at the classical level. The divergence of the supplementary term vanishes identically, so that the redefinition \eqref{eq:T_mu_nu_redefinition} needed to ensure conservation at the quantum level is not modified. For the classical energy density, this amounts to $T_{00}^{(0)} \to T_{00}^{(0)} - \xi \del_z^2 \phi_\kink^2$, while for the semiclassical correction, we have
\begin{align}
 \bra{\Omega_\kink} T^{(2)}_{00}(\tilde \vp^{(0)}) \ket{\Omega_\kink} & \to \bra{\Omega_\kink} T^{(2)}_{00}(\tilde \vp^{(0)}) \ket{\Omega_\kink} - \xi \del_z^2 \vp^{2, \semiclass}_\kink, \\
 \bra{\Omega_\kink} T^{(1)}_{00}(\tilde \vp^{(1)}) \ket{\Omega_\kink} & \to \bra{\Omega_\kink} T^{(1)}_{00}(\tilde \vp^{(1)}) \ket{\Omega_\kink} - 2 \xi \del_z^2 \left( \bra{\Omega_\kink} \tilde \vp^{(1)} \ket{\Omega_\kink} \phi_\kink \right),
\end{align}
see also the discussion in Section IIIB of \cite{RebhanEtAl}. In particular, neither the classical energy nor its semiclassical correction is affected. Note however, that with a non-minimal coupling, the semiclassical energy density of the soliton would no longer be symmetric under $z \to - z$, which is related to the fact that a non-minimal coupling breaks the shift symmetry $\phi \to \phi + 2 \pi \frac{m}{\sqrt{\lambda}}$ of the sine-Gordon model.

\item It is instructive to consider the situation in which, instead of requiring the Hamiltonian energy density $\rho_\kink$ to vanish as $z \to \pm \infty$, we allow for a finite limit $\rho_\infty$, i.e., we choose a value of $\Lambda \neq \sqrt{2}(e^{\gamma} m)^{-1}$. The corresponding $\rho_\kink$ is obtained by adding $\frac{\mu^2(z)}{m^2} \rho_\infty$ to \eqref{eq:rho_semiclass}. Using the same renormalization condition, i.e., the same value of $\Lambda$ in \eqref{eq:H}, also for the expansion around the vacuum, the corresponding energy density is then constant, $\rho_\vac = \rho_\infty$ (we are thus investigating the effect of deviating from the renormalization prescription amounting to normal ordering in the vacuum background). The shift in the renormalization condition then amounts to $T^{(2)}_{00,\kink} \to T^{(2)}_{00,\kink} + \frac{\rho_\infty}{m^2} ( \mu^2 - m^2 )$, so that the semiclassical correction to the energy is modified as\footnote{Recall that $T^{(1)}_{00}$ does not contribute to the energy, as it is a total derivative, \cf \eqref{eq:T_00_1}.}
\beq
\label{eq:E_rho_infty}
 E^\semiclass_\kink \to E^\semiclass_\kink - 6 \frac{\rho_\infty}{m}.
\eeq
At first sight, it seems counterintuitive that the semiclassical correction to the kink mass should depend on a renormalization condition which fixes the vacuum energy density. However, the result makes perfect sense if one takes into account that the redefinition (finite renormalization) \eqref{eq:redef_p2}, which gives rise to a non-zero $\rho_\infty = \frac{1}{2} m^2 \alpha$ has to be accompanied by the redefinition $\vp^4 \to \vp^4 + 6 \alpha \vp^2$, as a consequence of the ``expansion'' (or ``field independence'') axiom for the definition of renormalized Wick and time-ordered products \cite{HollandsWaldWick}.\footnote{This requirement states that renormalization should commute with functional differentiation \wrt $\vp$. It can be seen as the analog of the possibility to integrate by parts in the path integral approach.} Hence, such a redefinition amounts to a finite mass renormalization $m^2 \to m^2 - 6 \lambda \alpha$ in the original Lagrangian \eqref{eq:L}. But implementing such a shift of the mass in the classical contribution to the energy, one finds, with \eqref{eq:E_class}, that
\beq
 - 6 \lambda \alpha \frac{\del}{\del m^2} E^\class_\kink = - 3 m \alpha,
\eeq
which exactly coincides with the supplementary term on the \rhs of \eqref{eq:E_rho_infty}. One can easily convince oneself that this is not a coincidence.\footnote{The variation of $E^\class_\kink$ \wrt $m^2$ is only due the variation of the parameter $m^2$ in the Hamiltonian, i.e., the variation of the classical kink solution does not contribute, as it extremizes the energy.} We thus arrive at a coherent picture where deviations from the ``normal ordering'' renormalization prescription in the vacuum background are tantamount to a change of the mass $m^2$ in the original Lagrangian.

\item 
As explained in the Introduction, we think that a global renormalization, as perfomed in \cite{DashenEtAl, Dashen:1975hd} to arrive at \eqref{eq:E_Dashen}, is not entirely satisfactory from a conceptual point of view. It would thus be nice to derive the calculational rules used in \cite{DashenEtAl, Dashen:1975hd} from the locally covariant renormalization framework, which is, in our view, more fundamental. However, we presently do not see that this is possible.

\item A crucial step in our calculations was the redefinition \eqref{eq:T_mu_nu_redefinition} of the stress tensor. The problem with such a direct redefinition is that it is in general not clear whether the conservation of the stress tensor can be upheld at higher orders in the interactions, in contrast to the situation in which a conserved stress tensor is achieved by redefining the Wick powers as in \eqref{eq:redef_p2}, \eqref{eq:redef_del_p2} \cite{HollandsWaldStress}. 
It was recently shown \cite{CadamuroFrob} that for the case of the sine-Gordon model, a conserved stress tensor can be achieved also in the (non-perturbatively) interacting theory, by a modification which reduces to \eqref{eq:T_mu_nu_redefinition} in the semiclassical limit. It would be important to see whether a similar result also holds for $\phi^4$ theory.

\item As already mentioned, the direct redefinition \eqref{eq:T_mu_nu_redefinition} of the stress tensor is not necessary in higher dimensions, or for Dirac fields. It would thus be interesting to increase the number of dimensions, i.e., consider a string in $2+1$ or a domain wall in $3+1$ dimensions, or to investigate fermions in the kink background, and to again compare with results obtained in dimensional regularization \cite{RebhanEtAl} (for higher dimensions) or ``local mode regularization'' \cite{GoldhaberEtAl} (for fermions), or with results for the supersymmetric case \cite{Yamagishi, ShifmanEtAl} (which include fermions).

\end{itemize}

\subsection*{Acknowledgements}

J.Z.~would like to thank G.~Bergner for pointing out the literature on semiclassical corrections in soliton backgrounds. 
We would also like to thank an anonymous referee for pointing out to us the previous results obtained in \cite{GoldhaberEtAl} for the $\phi^4$ kink.
This work has been funded by the Deutsche Forschungsgemeinschaft (DFG) under Grant No. 406116891 within the Research Training Group RTG 2522/1.


\begin{thebibliography}{99}

\bibitem{SuSchriefferHeeger1979}
W.~P.~Su, J.~R.~Schrieffer and A.~J.~Heeger,
``Solitons in polyacetylene,''
Phys. Rev. Lett. \textbf{42}, 1698-1701 (1979)
doi:10.1103/PhysRevLett.42.1698

\bibitem{DashenEtAl}
R.~F.~Dashen, B.~Hasslacher and A.~Neveu,
``Nonperturbative Methods and Extended Hadron Models in Field Theory 2. Two-Dimensional Models and Extended Hadrons,''
Phys. Rev. D \textbf{10}, 4130-4138 (1974)
doi:10.1103/PhysRevD.10.4130

\bibitem{Coser:2014lla}
A.~Coser, M.~Beria, G.~P.~Brandino, R.~M.~Konik and G.~Mussardo,
``Truncated Conformal Space Approach for 2D Landau-Ginzburg Theories,''
J. Stat. Mech. \textbf{1412}, P12010 (2014)
doi:10.1088/1742-5468/2014/12/P12010
[arXiv:1409.1494 [hep-th]].

\bibitem{RychkovVitale}
S.~Rychkov and L.~G.~Vitale,
``Hamiltonian truncation study of the $\phi^4$ theory in two dimensions. II. The $\mathbb Z_2$ -broken phase and the Chang duality,''
Phys. Rev. D \textbf{93}, no.6, 065014 (2016)
doi:10.1103/PhysRevD.93.065014
[arXiv:1512.00493 [hep-th]].

\bibitem{Bajnok:2015bgw}
Z.~Bajnok and M.~Lajer,
``Truncated Hilbert space approach to the 2d $\phi^{4}$ theory,''
JHEP \textbf{10}, 050 (2016)
doi:10.1007/JHEP10(2016)050
[arXiv:1512.06901 [hep-th]].

\bibitem{Elias-Miro:2017tup}
J.~Elias-Miro, S.~Rychkov and L.~G.~Vitale,
``NLO Renormalization in the Hamiltonian Truncation,''
Phys. Rev. D \textbf{96}, no.6, 065024 (2017)
doi:10.1103/PhysRevD.96.065024
[arXiv:1706.09929 [hep-th]].

\bibitem{Serone:2019szm}
M.~Serone, G.~Spada and G.~Villadoro,
``$\lambda \phi_2^4$ theory \textemdash{} Part II. the broken phase beyond NNNN(NNNN)LO,''
JHEP \textbf{05}, 047 (2019)
doi:10.1007/JHEP05(2019)047
[arXiv:1901.05023 [hep-th]].

\bibitem{Chang:1976ek}
S.~J.~Chang,
``The Existence of a Second Order Phase Transition in the Two-Dimensional phi**4 Field Theory,''
Phys. Rev. D \textbf{13}, 2778 (1976)
[erratum: Phys. Rev. D \textbf{16}, 1979 (1977)]
doi:10.1103/PhysRevD.13.2778

\bibitem{Dashen:1975hd}
R.~F.~Dashen, B.~Hasslacher and A.~Neveu,
``The Particle Spectrum in Model Field Theories from Semiclassical Functional Integral Techniques,''
Phys. Rev. D \textbf{11}, 3424 (1975)
doi:10.1103/PhysRevD.11.3424

\bibitem{Coleman:1974bu}
S.~R.~Coleman,
``The Quantum Sine-Gordon Equation as the Massive Thirring Model,''
Phys. Rev. D \textbf{11}, 2088 (1975)
doi:10.1103/PhysRevD.11.2088

\bibitem{Weinberg}
E.J.~Weinberg,
``Classical Solutions in Quantum Field Theory,''
Cambridge University Press (2012).

\bibitem{HollandsWaldWick}
S.~Hollands and R.~M.~Wald,
``Local Wick polynomials and time ordered products of quantum fields in curved space-time,''
Commun. Math. Phys. \textbf{223}, 289-326 (2001)
doi:10.1007/s002200100540
[arXiv:gr-qc/0103074 [gr-qc]].

\bibitem{HollandsWaldStress}
S.~Hollands and R.~M.~Wald,
``Conservation of the stress tensor in interacting quantum field theory in curved spacetimes,''
Rev. Math. Phys. \textbf{17}, 227-312 (2005)
doi:10.1142/S0129055X05002340
[arXiv:gr-qc/0404074 [gr-qc]].

\bibitem{HollandsWaldReview}
S.~Hollands and R.~M.~Wald,
``Quantum fields in curved spacetime,''
Phys. Rept. \textbf{574}, 1-35 (2015)
doi:10.1016/j.physrep.2015.02.001
[arXiv:1401.2026 [gr-qc]].

\bibitem{Dirac34}
P.~Dirac,
``Discussion of the infinite distributions of electrons in the theory of the positron,'' Proc.\ Camb.\ Phil.\ Soc. 30 (1934) 150.

\bibitem{SchlemmerZahn}
J.~Schlemmer and J.~Zahn,
``The current density in quantum electrodynamics in external potentials,''
Annals Phys. \textbf{359}, 31-45 (2015)
doi:10.1016/j.aop.2015.04.006
[arXiv:1501.05912 [hep-th]].

\bibitem{Fulling}
S.~A.~Fulling,
``Aspects of Quantum Field Theory in Curved Space-time,''
Cambridge University Press (1989).

\bibitem{Milton}
K.~A.~Milton,
``The Casimir Effect,''
World Scientific (2001).

\bibitem{WernerssonZahn}
J.~Wernersson and J.~Zahn,
``Vacuum polarization near boundaries,''
Phys. Rev. D \textbf{103}, no.1, 016012 (2021)
doi:10.1103/PhysRevD.103.016012
[arXiv:2010.05499 [hep-th]].

\bibitem{OpenStrings}
J.~Zahn,
``Semiclassical energy of open Nambu-Goto strings,''
Phys. Rev. D \textbf{97}, no.6, 066028 (2018)
doi:10.1103/PhysRevD.97.066028
[arXiv:1605.07928 [hep-th]].

\bibitem{ClosedStrings}
M.~Kozo\v{n} and J.~Zahn,
``Semiclassical energy of closed Nambu-Goto strings,''
Phys. Rev. D \textbf{100}, no.10, 106005 (2019)
doi:10.1103/PhysRevD.100.106005
[arXiv:1610.02813 [hep-th]].

\bibitem{HellermanSwanson}
S.~Hellerman and I.~Swanson,
``String Theory of the Regge Intercept,''
Phys. Rev. Lett. \textbf{114}, no.11, 111601 (2015)
doi:10.1103/PhysRevLett.114.111601
[arXiv:1312.0999 [hep-th]].

\bibitem{GrahamEtAl}
N.~Graham, R.~L.~Jaffe, V.~Khemani, M.~Quandt, M.~Scandurra and H.~Weigel,
``Calculating vacuum energies in renormalizable quantum field theories: A New approach to the Casimir problem,''
Nucl. Phys. B \textbf{645}, 49-84 (2002)
doi:10.1016/S0550-3213(02)00823-4
[arXiv:hep-th/0207120 [hep-th]].

\bibitem{FermiPizzocchero}
D.~Fermi and L.~Pizzocchero,
``Local zeta regularization and the Casimir effect,''
Prog. Theor. Phys. \textbf{126}, 419-434 (2011)
doi:10.1143/PTP.126.419
[arXiv:1104.4330 [math-ph]].

\bibitem{GoldhaberEtAl}
A.~S.~Goldhaber, A.~Litvintsev and P.~van Nieuwenhuizen,
``Local Casimir energy for solitons,''
Phys. Rev. D \textbf{67}, 105021 (2003)
doi:10.1103/PhysRevD.67.105021
[arXiv:hep-th/0109110 [hep-th]].

\bibitem{RebhanEtAl}
A.~Rebhan, A.~Schmitt and P.~van Nieuwenhuizen,
``One-loop results for kink and domain wall profiles at zero and finite temperature,''
Phys. Rev. D \textbf{80}, 045012 (2009)
doi:10.1103/PhysRevD.80.045012
[arXiv:0903.5242 [hep-th]].

\bibitem{LocCovDirac}
J.~Zahn,
``The renormalized locally covariant Dirac field,''
Rev. Math. Phys. \textbf{26}, no.1, 1330012 (2014)
doi:10.1142/S0129055X13300124
[arXiv:1210.4031 [math-ph]].

\bibitem{MorettiStress}
V.~Moretti,
``Comments on the stress energy tensor operator in curved space-time,''
Commun. Math. Phys. \textbf{232}, 189-221 (2003)
doi:10.1007/s00220-002-0702-7
[arXiv:gr-qc/0109048 [gr-qc]].

\bibitem{ShifmanEtAl}
M.~A.~Shifman, A.~I.~Vainshtein and M.~B.~Voloshin,
``Anomaly and quantum corrections to solitons in two-dimensional theories with minimal supersymmetry,''
Phys. Rev. D \textbf{59}, 045016 (1999)
doi:10.1103/PhysRevD.59.045016
[arXiv:hep-th/9810068 [hep-th]].

\bibitem{Rajaraman}
R.~Rajaraman,
``Solitons and Instantons,''
North Holland (1982).

\bibitem{MatsumotoEtAl1979}
H.~Matsumoto, G.~Oberlechner, M.~Umezawa and H.~Umezawa,
``Quantum Soliton and Classical Soliton,''
J. Math. Phys. \textbf{20}, 2088 (1979)
doi:10.1063/1.523977

\bibitem{Gervais:1975pa}
J.~L.~Gervais, A.~Jevicki and B.~Sakita,
``Perturbation Expansion Around Extended Particle States in Quantum Field Theory. 1.,''
Phys. Rev. D \textbf{12}, 1038 (1975)
doi:10.1103/PhysRevD.12.1038

\bibitem{DecaniniFolacci}
Y.~Decanini and A.~Folacci,
``Hadamard renormalization of the stress-energy tensor for a quantized scalar field in a general spacetime of arbitrary dimension,''
Phys. Rev. D \textbf{78}, 044025 (2008)
doi:10.1103/PhysRevD.78.044025
[arXiv:gr-qc/0512118 [gr-qc]].

\bibitem{CadamuroFrob}
M.~B.~Fr\"ob and D.~Cadamuro,
``Local operators in the Sine-Gordon model: $\partial_\mu \phi \, \partial_\nu \phi$ and the stress tensor,''
[arXiv:2205.09223 [math-ph]].

\bibitem{Fewster:1998pu}
C.~J.~Fewster and S.~P.~Eveson,
``Bounds on negative energy densities in flat space-time,''
Phys. Rev. D \textbf{58}, 084010 (1998)
doi:10.1103/PhysRevD.58.084010
[arXiv:gr-qc/9805024 [gr-qc]].

\bibitem{Mukhopadhyay:2021wmu}
M.~Mukhopadhyay, E.~I.~Sfakianakis, T.~Vachaspati and G.~Zahariade,
``Kink-antikink scattering in a quantum vacuum,''
JHEP \textbf{04}, 118 (2022)
doi:10.1007/JHEP04(2022)118
[arXiv:2110.08277 [hep-th]].

\bibitem{Baacke}
J.~Baacke and N.~Kevlishvili,
``One-loop corrections to the string tension of the vortex in the Abelian Higgs model,''
Phys. Rev. D \textbf{78}, 085008 (2008)
[erratum: Phys. Rev. D \textbf{82}, 129905 (2010)]
doi:10.1103/PhysRevD.78.085008
[arXiv:0806.4349 [hep-th]].

\bibitem{Rebhan:1997iv}
A.~Rebhan and P.~van Nieuwenhuizen,
``No saturation of the quantum Bogomolnyi bound by two-dimensional supersymmetric solitons,''
Nucl. Phys. B \textbf{508}, 449-467 (1997)
doi:10.1016/S0550-3213(97)00625-1
[arXiv:hep-th/9707163 [hep-th]].

\bibitem{Yamagishi}
H.~Yamagishi,
``Soliton Mass Distributions in (1+1)-dimensional Supersymmetric Theories,''
Phys. Lett. B \textbf{147}, 425-429 (1984)
doi:10.1016/0370-2693(84)91396-0

\end{thebibliography}
\end{document}